\documentclass[12pt]{article}
\usepackage{amsfonts}
\usepackage{psfig}
\newcommand{\beq}{\begin{equation}}
\newcommand{\eeq}{\end{equation}}
\newcommand{\beqn}{\begin{eqnarray}}
\newcommand{\eeqn}{\end{eqnarray}}
\newcommand{\bea}[1]{\beq\begin{array}{#1}}
\newcommand{\eea}{\end{array}\eeq}

\newcommand{\summ}[2]{\sum\limits_{#1}^{#2}}

\newcommand{\Tr}[1]{\;{1\over #1}\mathop{\rm Tr}}

\newcommand{\ket}[1]{|\,#1\,\rangle}
\newcommand{\bra}[1]{\langle\,#1\,|}

\newcommand{\diff}{\partial}
\newcommand{\cC}{{\cal C}}

\newcommand{\NP}[3]{{\it Nucl. Phys. }{\bf #1} (#2) #3}

\newcommand{\PL}[3]{{\it Phys. Lett. }{\bf #1} (#2) #3}

\newcommand{\PR}[3]{{\it Phys. Rev. }{\bf #1} (#2) #3}

\newcommand{\CMP}[3]{{\it Comm. Math. Phys. }{\bf #1} (#2) #3}

\begin{document}
\date{}
\title{Gauge Invariant Monopoles\\
in Lattice SU(2) Gluodynamics.
\vskip-40mm
\rightline{\small ITEP-LAT-2002-04}
\rightline{\small MPI-PhT-2002-10}
\vskip 40mm
}
\author{F.V.~Gubarev\\
\\
{\small\it Institute of Theoretical and  Experimental Physics,}\\
{\small\it B.Cheremushkinskaya 25, Moscow, 117259, Russia}\\
}
\maketitle
\thispagestyle{empty}
\setcounter{page}{0}
\begin{abstract}
\noindent
We consider lattice implementation of the recently proposed 
gauge invariant definition of the monopole charge.
Because of the lattice discretization the algorithm gives rise to 
specific lattice artifacts and an effective Ising model.
The Ising-model problem might in principle be solved and we discuss
the role of the Maximal Abelian gauge in this respect.
The lattice artifacts are much more difficult to deal with since
they are mixed up with monopoles thus obscuring the physical observables.
Nevertheless, it is possible to extract the density of physical monopoles
which seems to scale correctly towards the continuum limit.
\end{abstract}

\newpage
%====================================================================
\subsection*{Introduction}
\noindent
In this paper we implement the gauge invariant monopole construction~\cite{main}
on the lattice. Since we use heavily the results of Ref.~\cite{main}
let us recall first  the essentials of the monopole definition in the continuum limit.

One starts with consideration of the fundamental Wilson loop $W(T)$,
$\Tr{2} W(T) = \cos\varphi(T)$
calculated on a closed contour $\cC(t)$, $t\in[0;T]$.
For a given $W(t)$ it is always possible to find a family of spin coherent
states $\ket{\vec{n}(t)}$ such that
\beq
\label{family}
e^{i\varphi(t)} \, \ket{\vec{n}(t)} ~=~ W(t)\,\ket{\vec{n}(0)}\,,
\qquad
\vec{n}(T) ~=~ \vec{n}(0)\,.
\eeq
Eq.~(\ref{family}) naturally assigns the unique (up to the sign, see below)
vector $\vec{n}(t)$ to each point on the contour
$\cC$. In terms of $\vec{n}(t)$ the phase angle $\varphi(T)$ of the Wilson loop
can be expressed as
\beq
\label{phase}
\varphi(T) ~=~ \frac{1}{2}\int\limits_\cC \vec{A}\vec{n}
~+~ \frac{1}{4} \int\limits_{S_\cC} \vec{n}\cdot\diff\vec{n}\times\diff\vec{n}\,,
\eeq
where $S_\cC$ is an arbitrary surface spanned on $\cC$ and $\vec{A}$ denotes
the tangential to $\cC$ components of the gauge potentials,
$A^a(t)~=~ A^a_\mu(x(t)) \;\dot{x}_\mu(t)$.
As is clear from Eq.~(\ref{family}), the initial state
$\ket{\vec{n}(0)}$ is an eigenstate of the full Wilson loop.  
Since $W(T)$ has two eigenstates
$\ket{\pm\vec{n}(0)}$ ('spin up' and 'spin down'), with the corresponding
families $\pm\vec{n}(t)$ and phases $\pm\varphi(T)$,
there is
a sign ambiguity in Eq.~(\ref{family}).

One considers next an arbitrary smooth closed surface $S^2_{phys}$ and
covers it with a set of infinitesimal patches of area
$\delta\sigma_x$. The phase evaluated on each patch is:
\beq
\label{small-phase}
\delta\varphi_x ~=~ \frac{1}{4}\,\{\, \diff\wedge(\vec{n}\vec{A}) ~+~ 
\vec{n}\cdot\diff\vec{n}\times\diff\vec{n}\,\} \,\, \delta\sigma_x\,.
\eeq
One can show that the usual assumption of the gauge fields continuity
implies existence of a smooth field $\vec{n}_x\in S^2_{phys}$.
Therefore it is legitimate to integrate Eq.~(\ref{small-phase}) over the 
surface $S^2_{phys}$:
\beq
\label{charge}
q ~=~ \frac{1}{2\pi} \oint\limits_{S^2_{phys}} \delta\varphi  ~=~
\frac{1}{8\pi} \oint\limits_{S^2_{phys}} 
\{\, \diff\wedge(\vec{n}\vec{A}) ~+~ \vec{n}\cdot\diff\vec{n}\times\diff\vec{n}\,\} \,\, \delta\sigma\,,
\eeq
and get a gauge invariant definition of the
monopole charge contained inside $S^2_{phys}$.
Note that for a smooth $S^2_{phys}$ and for continuous gauge fields the sign
ambiguity mentioned above is not important. In fact, it reduces to the
freedom to globally change the sign of $\vec{n}\in S^2_{phys}$
which in turn is equivalent to changing the sign of $q$, see Eq.~(\ref{charge}).

One can show~\cite{main} that at the classical level 
it is the singular Wu-Yang monopole solution~\cite{WuYa}
which corresponds to a non-trivial $q$.
In other words, Eq.~(\ref{charge}) provides us with
a gauge invariant topological definition of the
Wu-Yang monopole which can be used at the quantum level as well.
Moreover, the quantum generalization of the Wu-Yang monopole 
may look quite different from its classical counterpart.
Indeed, the classical
Wu-Yang solution is unstable~\cite{instability}.
Thus, the semi-classical approximation would be inconsistent. 
However, the instability of the Wu-Yang solution implies only that monopoles
(\ref{charge}) are irrelevant at short distances $r\ll\Lambda_{QCD}$.
As far as the infrared region is concerned  the dynamics
of  the monopoles (\ref{charge}) cannot be treated analytically
(see also Ref.~\cite{towards}).
One concludes, therefore, that at the quantum level the 
field configurations with a non-trivial charge $q$ have
little to do with the classical Wu-Yang solution both at short and large distances.
Moreover, there seem to be no analytical tools to investigate their dynamics.

Below we discuss the monopole definition (\ref{charge}) on the lattice.
Lattice implementation of the continuum ideas
is not always straightforward. A particular example is provided by instantons:
while in the continuum limit instantons are relatively simple objects,
their lattice definition is technically much more involved~\cite{Stone}.
Still, in principle, one may try to interpolate the lattice gauge fields
and apply the continuum ideas to the interpolated gauge potentials.
A drawback of this approach is that the interpolation is always not unique.

Because of the lattice discretization one cannot completely avoid 
interpolations while implementing Eq.~(\ref{charge}) on the lattice.
However, our interpolation procedure is unique and not applied to the
gauge fields themselves. Instead
we interpolate the states $\ket{\vec{n}}$ in between neighboring
lattice cells thus getting a continuous field $\vec{n}$ as required by
Eq.~(\ref{charge}). This enables us to construct integer valued conserved
monopole currents on the  lattice.

There are two disadvantages of our approach both of which we
discuss in detail later:

{\it i)} The sign ambiguity mentioned above which is harmless in the continuum
reemerges on the lattice in the form of an effective Ising model.
In turn, the problem of finding the gauge invariant monopoles on the lattice becomes
equivalent to finding the global minima of the corresponding Ising action,
which is non-trivial because the model is frustrated.
What is surprising, however, is that in the Maximal
Abelian gauge the ground state of the Ising model is approximated very well,
although we do not fix any particular gauge neither in our construction
nor in the numerical simulations.

{\it ii)} The interpolation of spin states in between neighboring lattice cells
enforces a change of the lattice geometry. Namely, one has to consider
additional three dimensional cells on the lattice which can contain gauge
invariant monopole charges. We argue below that these charges are
pure lattice artifacts which disappear in the limit of infinitesimal 
lattice spacing. The artifact's contribution to the physical
observables must be carefully subtracted which is a non-trivial problem
because of complicated lattice geometry and monopole-artifact mixing.
Nevertheless, for simple quantities like monopole density the monopole-artifact
separation turns possible. When this is done, the density of physical monopoles
scales correctly towards the continuum limit.

%====================================================================
\subsection*{1. 'Stokes Theorem'}
\noindent
Complexity of the lattice implementation of Eq.~(\ref{charge})
might already be visualized from that equation itself. Indeed, the effective 'Higgs field'
$\vec{n}$ is unusual since it is defined not on space-time points (lattice sites)
but on elementary two-dimensional cells (plaquettes). Moreover, the field
$\vec{n}$ is defined ~\cite{main} in terms of the field strength
tensor $F^a_{\mu\nu}$ which itself
is badly defined on the lattice. Indeed,
in the continuum limit the Wilson loop calculated on a plaquette reduces to
a particular component of $F_{\mu\nu}$ regardless of the initial point from
which we started evaluation of Wilson loop.
But for a finite lattice spacing the Wilson loop  depends drastically
on the starting point. Because of this, the spin state which we would
like to assign to a given plaquette also depends on the plaquette definition.

\begin{figure}[ht]
\centerline{\psfig{file=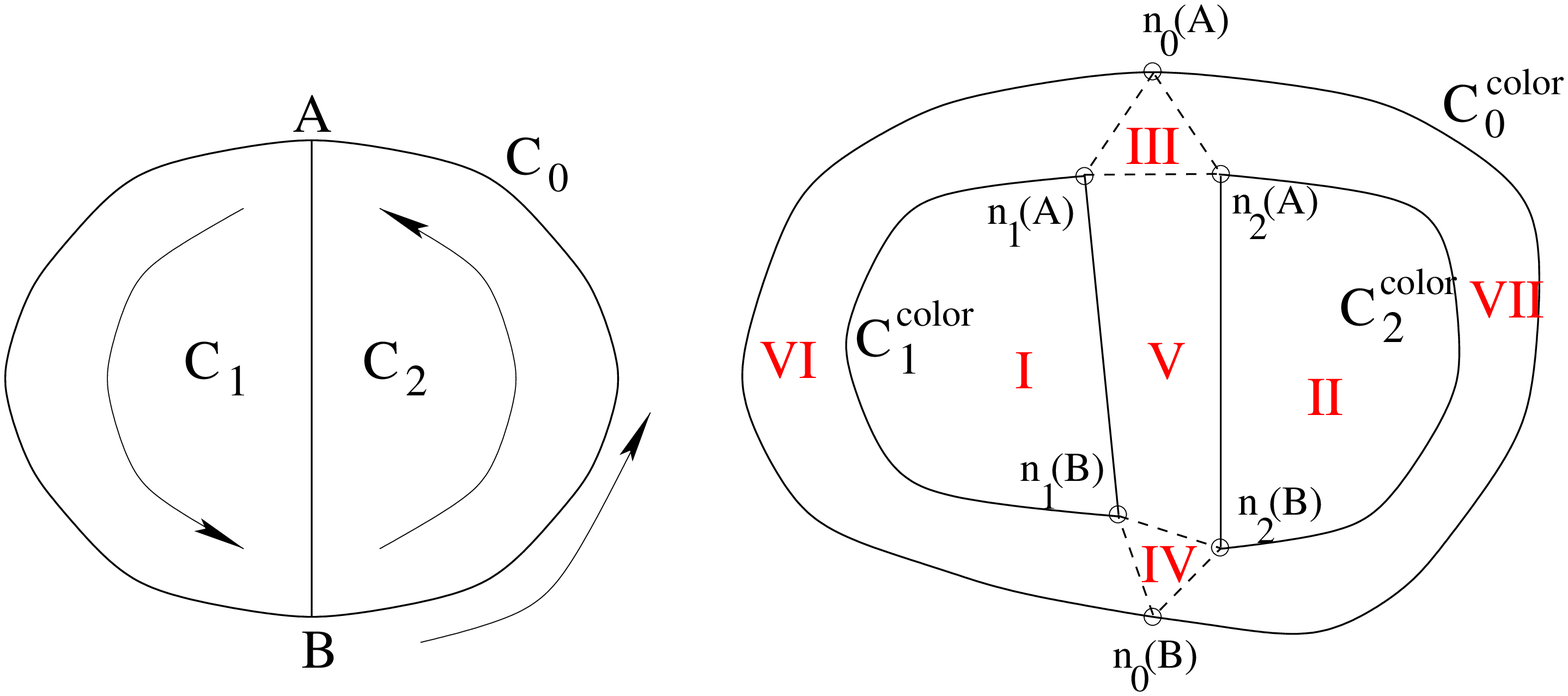,width=0.8\textwidth,silent=}}

\vspace{3mm}

\centerline{Fig.~1}
\end{figure}

The situation becomes even worse if we consider a common link of two plaquettes.
The spin states are to be constructed  on each plaquette separately and they
have nothing in common, in contrast with the continuum case.
To get insight into the problem, let us consider an arbitrary contour
$\cC_0$ and dissect it in points $A$ and $B$ into two smaller pieces
$\cC_1$ and $\cC_2$ (Fig.~1).
The spin states are introduced for each contour via Eq.~(\ref{family})
taking for definiteness the 'spin up' initial states. In this way one gets three different
families $\vec{n}_i(t)$ associated with each contour $\cC_i$. Suppose that we know the
phases $\varphi_i$ of the Wilson loops calculated on $\cC_i$. What is the relation between them?

The answer may be found by considering the image of each contour under the
map: $\vec{n}_i:\cC_i\to\cC^{color}_i$. Note that the two triples
$\{\vec{n}_i(A)\}$, $\{\vec{n}_i(B)\}$, $i=0,1,2$
are distinguishable since the relative position of the vectors in each triple
is gauge invariant.
Moreover, there exists a simple interpretation of
the original Wilson loops
in terms of $\cC^{color}_i$.
Namely, $W_i(t)$ describes periodic motion of a point-like
charged particle on $S^2_{color}$ since 
every closed path $C^{color}_i$ is associated with a unique phase factor.
Therefore, in terms of $C^{color}_i$ the problem becomes essentially Abelian
and one could have 
concluded that $\varphi_0$ is equal to the sum of phases calculated on 
the pre-images of the closed paths
I, II, ..., VII (see Fig.~1). But there is a loophole in this argumentation: we do not
know the $SU(2)$ matrices $g_{ij}$ which connect the points $\vec{n}_i\to\vec{n}_j$:
\beq
\label{connect}
g_{ij}(A):\,\vec{n}_i(A)\to\vec{n}_j(A)\,,
\qquad
g_{ij}(B):\,\vec{n}_i(B)\to\vec{n}_j(B)
\eeq
and which are denoted by dashed lines on the Fig.~1. Clearly, $g_{ij}$ might be an arbitrary
$SU(2)$ elements provided that Eq.~(\ref{connect}) holds true.
 
Among various $g\in SU(2)$ which connect two given points $\vec{n}_1, \vec{n}_2 \in S^2_{color}$,
$g\ket{\vec{n}_1}\sim\ket{\vec{n}_2}$ there is a special one. Namely, if $g$ describes
the motion $\vec{n}_1\to\vec{n}_2$ along the shortest geodesic connecting $\vec{n}_1$, $\vec{n}_2$
then $g$ is uniquely defined:
\beq
\label{geodesic}
g ~=~ (\vec{n}_1 \vec{n}_2) ~+~   i \sigma^a [\vec{n}_1\times\vec{n}_2]^a\,.
\eeq
We call it the geodesic matrix below. Moreover, for the geodesic matrices $g_{ij}$
the relation between the phases $\varphi_i$ becomes extremely simple. Indeed,
one can readily verify that in this case the phase angles associated with cells V, VI, VII
are vanishing while  for the the cells III, IV they are equal to the oriented
areas of the spherical triangles $\{\vec{n}_i(A)\}$ and $\{\vec{n}_i(B)\}$, respectively.
Thus, we get the following simple equation which relates the phases 
of the Wilson loops:
\beq
\label{stocks}
\varphi_0 ~=~  \varphi_1 ~+~ \varphi_2 ~+~ \gamma(A) ~+~ \gamma(B)\,,
\eeq
$$
\gamma(N)~=~ \,\mbox{ oriented solid angle between }\, \vec{n}_0(N), \vec{n}_1(N), \vec{n}_2(N)\,.
$$
Note that we always implicitly assume $\mbox{mod }2\pi$ operation~\cite{main}.

%====================================================================
\subsection*{2. Single Plaquette Construction.}
\noindent
We start the construction of monopoles (\ref{charge}) on the lattice following the same
strategy as in the continuum limit. Namely, the simplest object which we consider first
is an elementary plaquette, see Fig.~2.
\begin{figure}[ht]
\centerline{\psfig{file=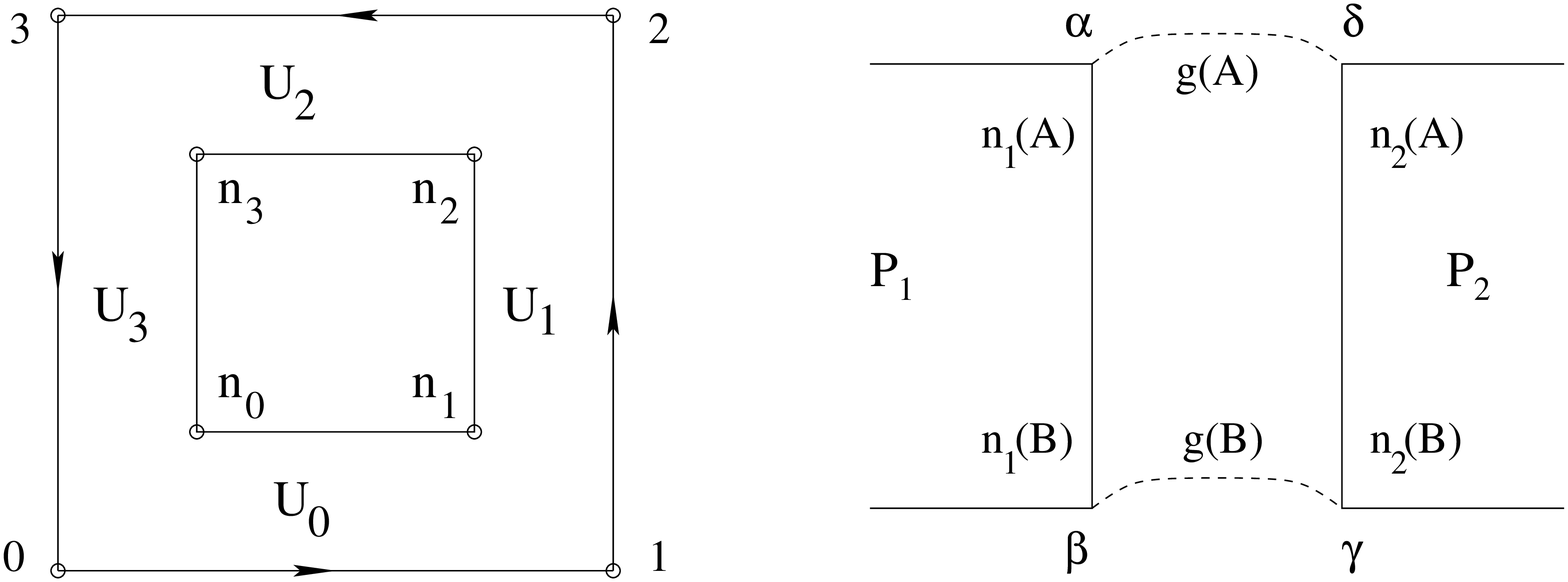,width=0.75\textwidth,silent=}}

\centerline{Fig.~2~~~~~~~~~~~~~~~~~~~~~~~~~~~~~~~~~~~~~~~~~~~~Fig.~3.}
\end{figure}

As is noted above for any finite lattice spacing the matrix $U_p = U_1U_2U_3U_4$
is to be treated as a large Wilson loop. In particular,
it is impossible to assign a single well defined spin state to a given plaquette.
On the other hand, the Eq.~(\ref{family}) is still valid and in fact directly
applicable. Namely, one can calculate the eigenvector of the matrix $U_p$:
\beq
\label{init}
\bra{\vec{n}_0}\,U_p ~=~ e^{i\varphi_p}\,\bra{\vec{n}_0}
\eeq
and then propagate the initial state $\bra{\vec{n}_0}$ along the plaquette's edges:
\beq
\label{latevolution}
\bra{\vec{n}_k}\,U_k ~=~ e^{i\varphi_k}\,\bra{\vec{n}_{k+1}}\,,
\qquad k = 0,...,3\,, \qquad \bra{\vec{n}_4} ~=~ \bra{\vec{n}_0} \,.
\eeq
In this way one gets four states $\bra{\vec{n}_k}$ sitting at the plaquette's corners.
Moreover, each
$\bra{\vec{n}_k}$ is an eigenstate of the plaquette matrix calculated starting from point $k$, e.g.
$\bra{\vec{n}_1} U_1U_2U_3U_0 \sim \bra{\vec{n}_1}$ etc.
Therefore, the vectors $\vec{n}_k$ are in a sense always parallel to the color
magnetic field piercing the plaquette considered.
Moreover, Eqs.~(\ref{init},\ref{latevolution}) imply the
following simple relation between the plaquette
angle $\varphi_p$ and phases $\varphi_i$ associated with the corresponding links:
\beq
\varphi_p ~=~ \summ{k=0}{3} \varphi_k\,.
\eeq

Note that we still have a freedom to choose an arbitrary sign of the initial state
(\ref{init}), the same as in the continuum limit. It is clear that the
change $\vec{n}_0\to -\vec{n}_0$ would also
flip the signs of all other states: $\vec{n}_k\to -\vec{n}_k$. Thus, it makes sense
to speak about the sign associated with each plaquette.
While only a single plaquette is considered the sign ambiguity is in fact irresolvable. 
Note also that for pure diagonal link matrices all vectors $\vec{n}_k$ collapse
to a single point on the $S^2_{color}$.

Next, consider a common link of two adjacent plaquettes, see Fig.~3.
Since the spin states are constructed separately on each plaquette the states
$\vec{n}_1(A)$, $\vec{n}_2(A)$ and $\vec{n}_1(B)$, $\vec{n}_2(B)$
are quite different in general. It is important, however, that the products
$\vec{n}_1(A)\cdot\vec{n}_2(A)$ and $\vec{n}_1(B)\cdot\vec{n}_2(B)$ are gauge invariant.
Moreover, in case of diagonal lattice gauge fields
(or for a gauge copy of the diagonal configuration) we have
\beq
\label{abelian-spins}
\vec{n}_1(A)\cdot\vec{n}_2(A) ~=~ \vec{n}_1(B)\cdot\vec{n}_2(B) ~=~ \pm 1\,,
\eeq
depending on the relative sign of initial states chosen on each plaquette.
Therefore, there is a natural way to fix the relative sign on $P_1$ and $P_2$
by the requirement that across the common link of two plaquettes the spins should
be as smooth as possible. In fact, this approach is essentially the same as
in the continuum limit. The only difference is that in the continuum
we indeed do have a continuous distribution of spins while on the lattice the
true smoothness of $\vec{n}_x$ is not possible. The only exception is a particular
case $|\vec{n}_1(A)\cdot\vec{n}_2(A)| = 1$ for which everything is in fact trivial. 

Thus, on the lattice we inevitably have to interpolate the states $\vec{n}_1(A)$
and $\vec{n}_2(A)$ by connecting them with an appropriate $SU(2)$ matrix $g$ such that
$\bra{\vec{n}_1} g \sim \bra{\vec{n}_2}$. As we have argued in the previous section there is
a natural unique choice of $g$, namely the geodesic interpolation, Eq.~(\ref{geodesic}).
This choice is singled out, in particular, since only in this case we do not introduce
additional fluxes in the intersection of two plaquettes:  the Wilson loop calculated
on 2-cell $(\alpha\beta\gamma\delta)$ is exactly unity
no matter what the lattice gauge fields are.

%==========================================================================
\subsection*{3. Monopole Charge on 3D Cube.}
\noindent
The above construction allows to formulate the notion of the monopole charge for a single
three-dimensional lattice cube. Namely, one has to apply the procedure of the previous section
to all six plaquettes of a given cube, Fig.~4.

\begin{figure}[ht]
\centerline{\psfig{file=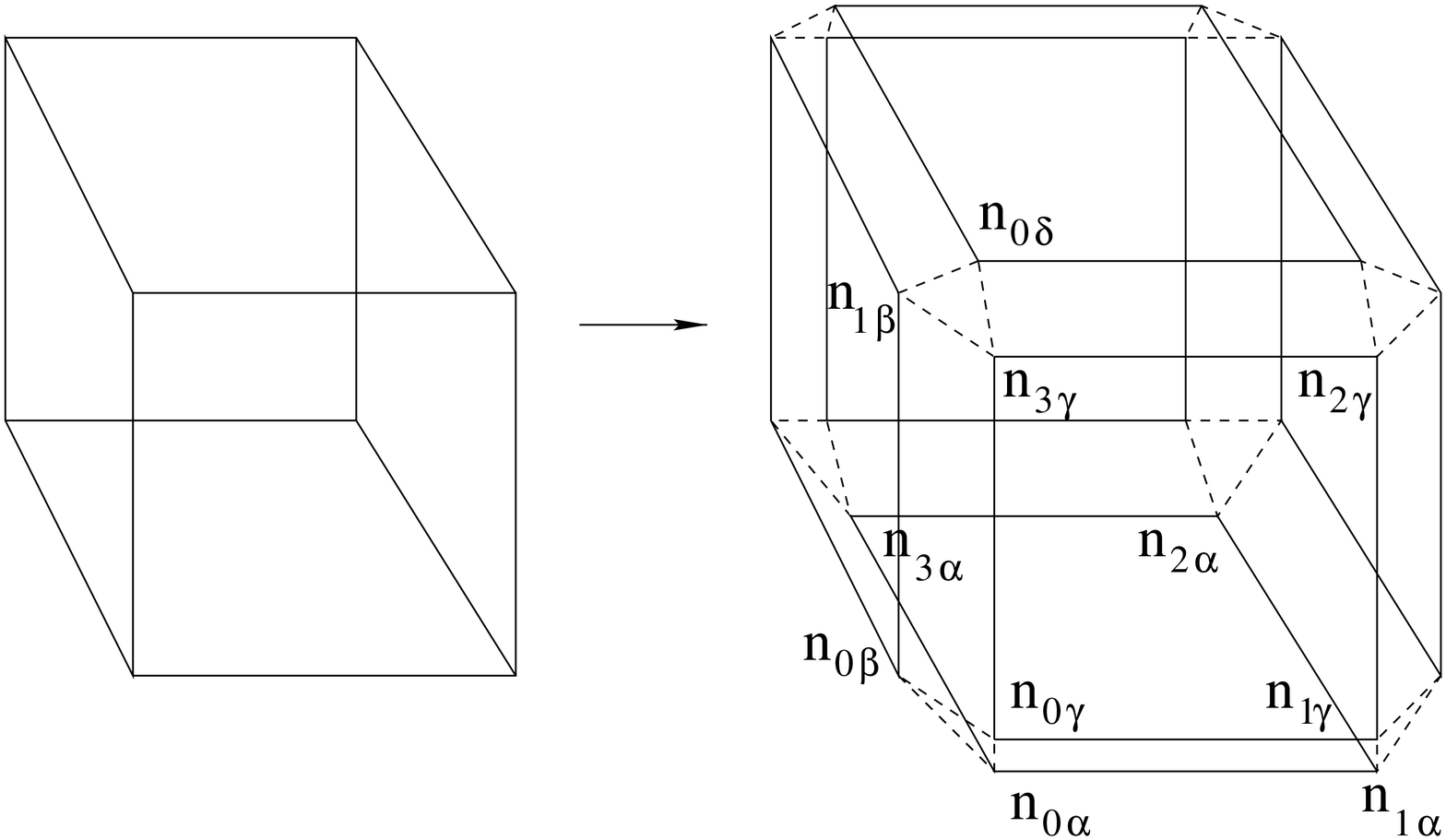,width=0.75\textwidth,silent=}}

\centerline{Fig.~5}
\end{figure}

Let us briefly summarize the essential steps of the construction.

{\it i)}  On each face $\alpha=0,...,5$ of the cube one calculates the corresponding
plaquette matrix $U_\alpha$ and then constructs four vectors $\vec{n}_{i,\alpha}$, $i=0,...,3$
via Eq.~(\ref{latevolution}) taking the initial state $\vec{n}_{0,\alpha}$ as an eigenstate
of $U_\alpha$ with arbitrary sign. 

{\it ii)} In order to get the smoothest distribution of $\vec{n}_{i,\alpha}$ on the cube boundary,
one considers the following Ising-like action:
\beq
\label{ising}
S ~=~ - \summ{\alpha\beta}{} z_\alpha z_\beta \, \summ{i,j}{} \vec{n}_{i,\alpha} \cdot \vec{n}_{j,\beta}\,,
\eeq
where dynamical variables are the signs $z_\alpha = \pm 1$ associated with each plaquette,
the first sum is over all neighboring 2-cells in the cube's boundary and the summation in
$i$, $j$ indexes extends over all neighboring spins. The action (\ref{ising}) is
minimized with respect
to all signs $z_\alpha$ and, finally, for negative $z_\alpha$ the corresponding
vectors $\vec{n}_{i,\alpha}$  are inverted
$\vec{n}_{i,\alpha} \to - \vec{n}_{i,\alpha}$. It is clear that the ground state of the
Ising model (\ref{ising}) corresponds to the smoothest possible distribution of
$\vec{n}_{i,\alpha}$. Moreover, for a single cube the ground state may be found exactly.

{\it iii)} Finally, the three vectors around each vertex of the cube are geodesically
interpolated (dashed lines on Fig.~5). Note that the relative position of vectors
in each triple, say
$(\vec{n}_{1,\beta}\,\vec{n}_{3,\gamma}\,\vec{n}_{0,\delta})$ is gauge invariant
and therefore the geodesic interpolation is a well defined prescription.

The magnetic charge contained in a given cube is proportional to the net outgoing
magnetic flux, cf. Eq.~(\ref{charge}).
Namely, one sums up the phases $\varphi_i$ associated via Eq.~(\ref{init})
with all two-dimensional cells on the boundary of the 'extended' cube (Fig.~5, right).
Note that due to the geodesic interpolation not all the 2-cells must be taken
into account. For example, the flux emanating through the 2-cell
$(\vec{n}_{1,\beta}, \vec{n}_{0,\beta}, \vec{n}_{0,\gamma}, \vec{n}_{3,\gamma})$
is exactly zero. The final expression for the magnetic charge is therefore
\beq
\label{latcharge}
q ~=~ \frac{1}{2\pi} \,
\left[ \summ{plaq~\alpha=0}{5} \varphi_\alpha ~+
\summ{vertex~v=0}{7} \tilde{\varphi}_v \, \right]\,,
\eeq
where $\varphi_\alpha$ is the plaquette angle (\ref{init}) and
$\tilde{\varphi}_v$ is the flux corresponding to the triangle (dashed lines)
at vertex $v$ which in turn is equal to the oriented solid angle between
vectors $\vec{n}_{i,\alpha}$ at the triangle's corners.

Eq.~(\ref{latcharge}) has in fact a transparent meaning. The first term counts
the flux emanating through the original lattice plaquettes and therefore it is
the physical flux of the monopole. The second term is in a sense a correction due to the
lattice coarseness. Indeed, there were two important requirements
in the corresponding continuum construction : $S^2_{phys}$ should be smooth
and $\delta\varphi_x$ is to be calculated on an infinitesimal contour $\delta\cC_x$.
In fact none of them is realized on the lattice since the lattice cube
for sure cannot be considered as a smooth surface and the lattice plaquette is not
infinitesimal in any sense. In fact the second term in Eq.~(\ref{latcharge})
just accounts for these facts by plugging the holes between the
neighboring vectors $\vec{n}_x$.
Let us emphasize that the appearance of the additional contribution,
Eq.~(\ref{latcharge}), is entirely due to the lattice discretization.
There are no similar term in the continuum expression (\ref{charge}).

However, there exists a particular case when the second contribution,
Eq.~(\ref{latcharge}), is zero. This happens when all the neighboring spins
$\vec{n}_{i,\alpha}$ are aligned with each other. From Eq.~(\ref{abelian-spins})
we conclude then that this corresponds to a pure Abelian configuration. Namely, there
exists a particular gauge in which all the lattice gauge fields are diagonal. 
Moreover, Eq.~(\ref{latcharge}) is then identical to the well known 
DeGrand--Toussaint definition of the Abelian monopole~\cite{deGrand}.

%==========================================================================
\subsection*{4. Effective Ising Model and Lattice Artifacts.}
\noindent
In this section we discuss some details of the above 
construction which are in fact crucial in the actual numerical implementations.
For simplicity, we consider in this section three dimensional lattice only.
Let us start from the Eq.~(\ref{ising}) which encodes the way how to get
the smoothest possible distribution of spins $\vec{n}_{i,\alpha}$.
In fact, it is quite important to find the global minimum of the action
(\ref{ising}). Indeed, let us consider an almost Abelian gauge fields
configuration for which 
\beq
n^a_{i,\alpha} ~\approx~ \pm \delta^{a3} \qquad \forall \, i, \alpha\,.
\eeq
Suppose that instead of the ground  state of the model (\ref{ising})
we are considering a local minimum of the action (\ref{ising}). After performing
the corresponding inversions $\vec{n}_{i,\alpha}\to -\vec{n}_{i,\alpha}$
for all the negative $z_\alpha$  we get a configuration of vectors $\vec{n}_{i,\alpha}$
almost all of which satisfy $\vec{n}_{i,\alpha} \approx \delta^{a3}$. But since
the minimum is only local there are some negative spins as well. Evidently,
Eq.~(\ref{latcharge}) when applied for such $\{\vec{n}_{i,\alpha}\}$
still produces a non zero charge which however has nothing to do with
reality since the contribution comes mostly from the second term of
Eq.~(\ref{latcharge}).

Therefore, it is indeed important to
find the true ground state of the model (\ref{ising}) before applying Eq.~(\ref{latcharge}).
One expects, however, that in the real simulations it is almost impossible to find
the global minima since for non-Abelian gauge fields the model
(\ref{ising}) is frustrated. Thus the concrete algorithm used to minimize
(\ref{ising}) might be crucial. 

As a matter of fact, the naive local minimization of (\ref{ising}) is doomed to produce
unphysical results for the monopoles (\ref{latcharge}). One may try to improve
the iterative minimization by preconditioning it with a particular gauge fixing
of the original lattice gauge fields. In our numerical tests (see below)
we have considered three particular gauge fixing prescriptions: 
minimal and center Landau and Maximal Abelian gauges which are defined via
maximization of the following functionals
\beq
\label{gauges}
\summ{l}{} \Tr{2} U_l
\,,\qquad
\summ{l}{} \left(\Tr{2} U_l\right)^2
\,,\qquad
\summ{l}{} \Tr{2}\,U_l\sigma^3 U^+_l\sigma^3~,
\eeq
respectively. Our finding is that the Maximal Abelian gauge is indeed greatly
favored\footnote{
%- - - - - - - - - - - - - - - - - - - - - - - - - - - - - - - - - - - - -
See the next section for details on numerics.
%- - - - - - - - - - - - - - - - - - - - - - - - - - - - - - - - - - - - -
}. Namely, when the lattice fields are first MAA gauge fixed then for each $\alpha$
it is sufficient to pick up the nearest to the north pole family
$\vec{n}_{i,\alpha}$. The resulting distribution $\{\vec{n}_{i,\alpha}\}$
almost exactly corresponds to the ground state of (\ref{ising}).
For other gauges the result is less pronounced: although Landau gauge fixing
improves the local minimization the resulting minimal value
of (\ref{ising}) is considerably larger than in case of the Maximal Abelian gauge.

Let us note that in our numerical simulations none of the gauges (\ref{gauges}) was used.
Instead, we have applied a variant of the simulated annealing algorithm~\cite{SA}
to directly minimize the action (\ref{ising}). As one may expect the latter approach
turns out to be superior to any kind of local minimization.

Let us turn now to another major problem of the lattice implementation of
Eq.~(\ref{charge}).  Namely, one can see that Fig.~5 implies in fact a change
of the lattice geometry since this construction being iterated for all 3-cubes
produces 'extended' cubes instead and introduces new three dimensional
cells on the lattice.  In particular, 
in the intersection of four cubes one gets the geometrical figure depicted on 
the Fig.~6
and a new 3-cell at each site of the original lattice, see Fig.~7.

\begin{figure}[ht]
\centerline{\psfig{file=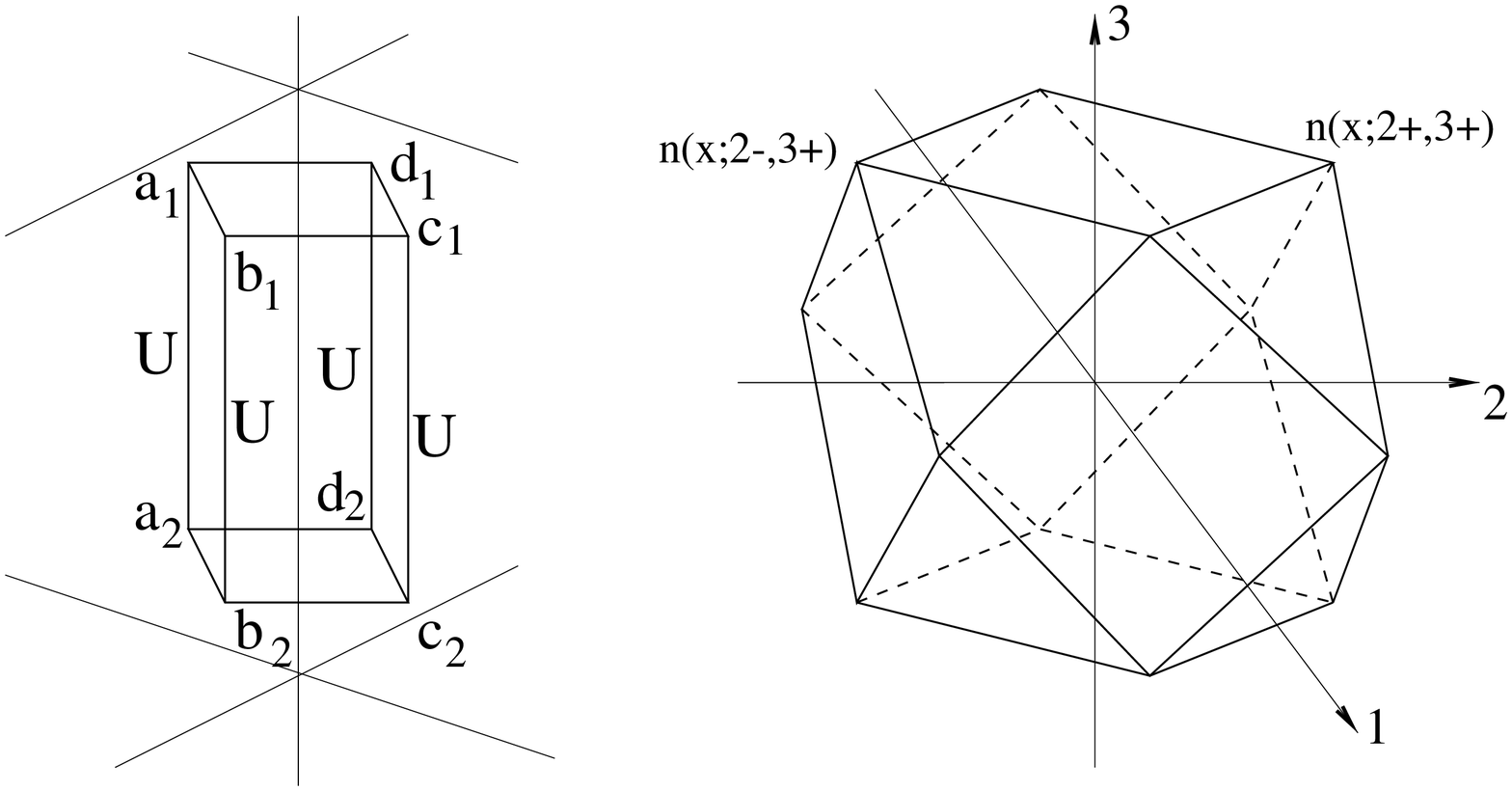,width=0.75\textwidth,silent=}}

\centerline{Fig.~6~~~~~~~~~~~~~~~~~~~~~~~~~~~~Fig.~7}
\end{figure}

Consider first the Fig.~6. For the plotted 3-cell the faces $(a_ib_ic_id_i)$, $i=1,2$
consist of the geodesic matrices while all others were in fact already discussed.
Namely we have shown that fluxes piercing the cells
$(b_1b_2c_2c_1)$, $(a_1,a_2,b_1,b_2)$ etc. are exactly zero. Moreover, since there is
the same matrix $U$ on each link $(a_1a_2)$, $(b_1b_2)$ etc. one concludes that 
the images of squares $(a_ib_ic_id_i)$, $i=1,2$ under the maps
$\vec{n}(a_i)$, $\vec{n}(b_i)$ etc. are isomorphic.
Due to the properties of the geodesic interpolation the fluxes piercing
$(a_1b_1c_1d_1)$ and $(a_2b_2c_2d_2)$ are exactly the same.

The overall conclusion about Fig.~6 is as follows. The magnetic charge inside
the plotted 3-cell is identical zero. Moreover, the flux which goes
from this cell to the neighboring cubes is zero either. 
The whole flux propagates unmodified
along the link of the original lattice from the face $(a_1b_1c_1d_1)$ to the face
$(a_2b_2c_2d_2)$, see Fig.~6.

Turn now to the 3-cell on the Fig.~7. Its properties are easy
to derive. Namely, at each vertex there is a corresponding vector $\vec{n}_i$,
$i=0,...,11$. It is convenient to parameterize the vertices as
$(x;\hat{\mu},d_\mu;\hat{\nu},d_\nu) \equiv (x;\hat{\nu},d_\nu;\hat{\mu},d_\mu)$,
where $\hat{\mu}$, $\hat{\nu}$ are the
unit vectors along the lattice axes and $d_\mu$, $d_\nu$ are the corresponding shift signs
(see Fig.~7). Moreover, the pairs of vertices 
$$
(x;\hat{\mu},d_\mu;\hat{\nu},d_\nu), \, (x;\hat{\mu},d_\mu;\hat{\lambda},d_\lambda)
\qquad
\forall\, \mu, \nu, \lambda
$$
are connected by a link which carries an $SU(2)$ matrix $g$ such that 
$$
g:\,\vec{n}(x;\hat{\mu},d_\mu;\hat{\nu},d_\nu)
\to\vec{n}(x;\hat{\mu},d_\mu;\hat{\lambda},d_\lambda)
$$
corresponds to the geodesic motion on $S^2_{color}$.
Note that the relative position of all $\vec{n}(x;\hat{\mu},d_\mu;\hat{\nu},d_\nu)$
is gauge invariant and therefore the construction of the 3-cell on the Fig.~7
is well defined.

From the above properties one concludes that fluxes emanating from the faces
of the presented 3-cell are equal to the oriented solid angles between the 
corresponding vectors situated at the vertices of that face. 
Then it follows immediately that the magnetic charge inside this 3-cell is
gauge invariant and in general  non-zero. Thus, we get another type of monopoles which however
live in the sites of original lattice contrary to the monopoles in Eq.~(\ref{latcharge}),
which belong to the dual lattice, as usual. In fact, there is
another striking property of the monopoles on the Fig.~7.
Namely, none of the fluxes which this monopole emanates enter the original
Wilson action of the lattice gauge models. Therefore, naively this monopole costs
no action at all.

One can readily convince oneself that the actionless monopoles
are of the same origin as the second term in Eq.~(\ref{latcharge}).
They appear because of the violation on the lattice
of the two major requirements of continuum
theory  (see the discussion after Eq.~(\ref{latcharge})).
We conclude, therefore, that the actionless monopoles are intrinsic
to the lattice formulation. They disappear only in the academic al limit
of such small lattice spacing that one can indeed identify the lattice
plaquette with a particular component of continuum $F_{\mu\nu}$. Of course, 
in this limit the above single cube construction is not applicable any longer
and one has to consider the extended monopoles~\cite{Extend}.

Thus, the actionless monopoles are in fact lattice artifacts 
which exist only because of the lattice discretization. When any physical
observable is measured on the lattice the contribution of this objects
must be carefully subtracted -- the task which is quite non-trivial by itself.
Indeed, even if $D=3$ and the lattice geometry is well understood (see Figs.~5-7)
there seems to be no unique way to extract, e.g., density of the physical
monopoles. In three dimensions the monopoles are point-like and one may
naively expect that each artifact should form a small dipole-like structure
either with 'physical' monopole or with another artifact. But this expectation
is in fact wrong since the flux emanating from each artifact is not suppressed
by the action. Therefore, e.g., artifact-artifact chain might be arbitrarily
prolongated. 

In four dimensions the situation is even worse because of 
the complicated lattice geometry.
The monopole currents are of course conserved, but the conservation takes place
not on the hypercubical lattice. As a result the monopoles may freely propagate
from 3-cubes to the sites and back on the original lattice. 
Thus, discrimination against the artifacts
becomes ambiguous. Nevertheless in the next section we show that in $D=4$ the simplest
'brute-force' approach works unexpectedly well for the monopole density.

We also would like to mention that there exists another view on the
monopole-artifact mixing. Indeed, suppose for a while that the second
term in Eq.~(\ref{latcharge}) is zero. Then the monopole and artifact
currents are separately conserved on the hypercubical dual and original
lattices, respectively,
 and therefore there is no mixing between them. Thus it is the second term in
Eq.~(\ref{latcharge}) which mixes together the two types of magnetic charges.

Finally, let us note that there might be another way to fight
against the artifacts on the lattice. Here we mean in particular a kind of
blocking procedure which preserves the infrared properties
of the theory while filtering out the ultraviolet noise. This might be helpful
in approaching the limit of infinitesimal lattice spacing where the artifacts for
sure disappear.

%--------------------------------------------------------------------
\subsection*{5. Results of Numerical Simulations.}
\noindent
In this section we describe the results of our numerical simulations of the four dimensional
pure $SU(2)$ lattice gauge model. We have noted already that it is in fact crucial
to have a good approximation to the ground state of the model (\ref{ising})
which amounts to the ability to approximate the global minima of the corresponding action.
We have considered three different types of minimization procedure, namely, the naive
local minimization with and without gauge fixing preconditioning and a variant of the
simulated annealing algorithm.

The gauges considered  were described above, 
see Eq.~(\ref{gauges}), and each gauge was fixed
by the iterative maximization of the corresponding functional.
The gauge was considered fixed when all gauge rotation matrices $\Omega_x$
become sufficiently small:
$1-\Tr{2}\Omega_x < 10^{-6}$. The simulated annealing algorithm which we used
is the following. The action (\ref{ising}) was multiplied by a factor $\gamma$
(coupling constant) and then the quantum dynamics of variables $\{z_\alpha\}$
was simulated with the standard Metropolis algorithm. From the initial value
$\gamma \approx 0.1$ the coupling constant was moved towards $\gamma=\infty$
with steps $\Delta\gamma = 0.1$. Fifteen thermalization steps
were performed for each $\gamma$  and the algorithm was stopped when
the acceptance rate became smaller than 0.5\%.
After that on all plaquettes with $z_\alpha = -1$ the corresponding
$\vec{n}_{i,\alpha}$, $i=0,...,3$ were inverted $\vec{n}_{i,\alpha}\to - \vec{n}_{i,\alpha}$.

\begin{figure}[ht]
\centerline{\psfig{file=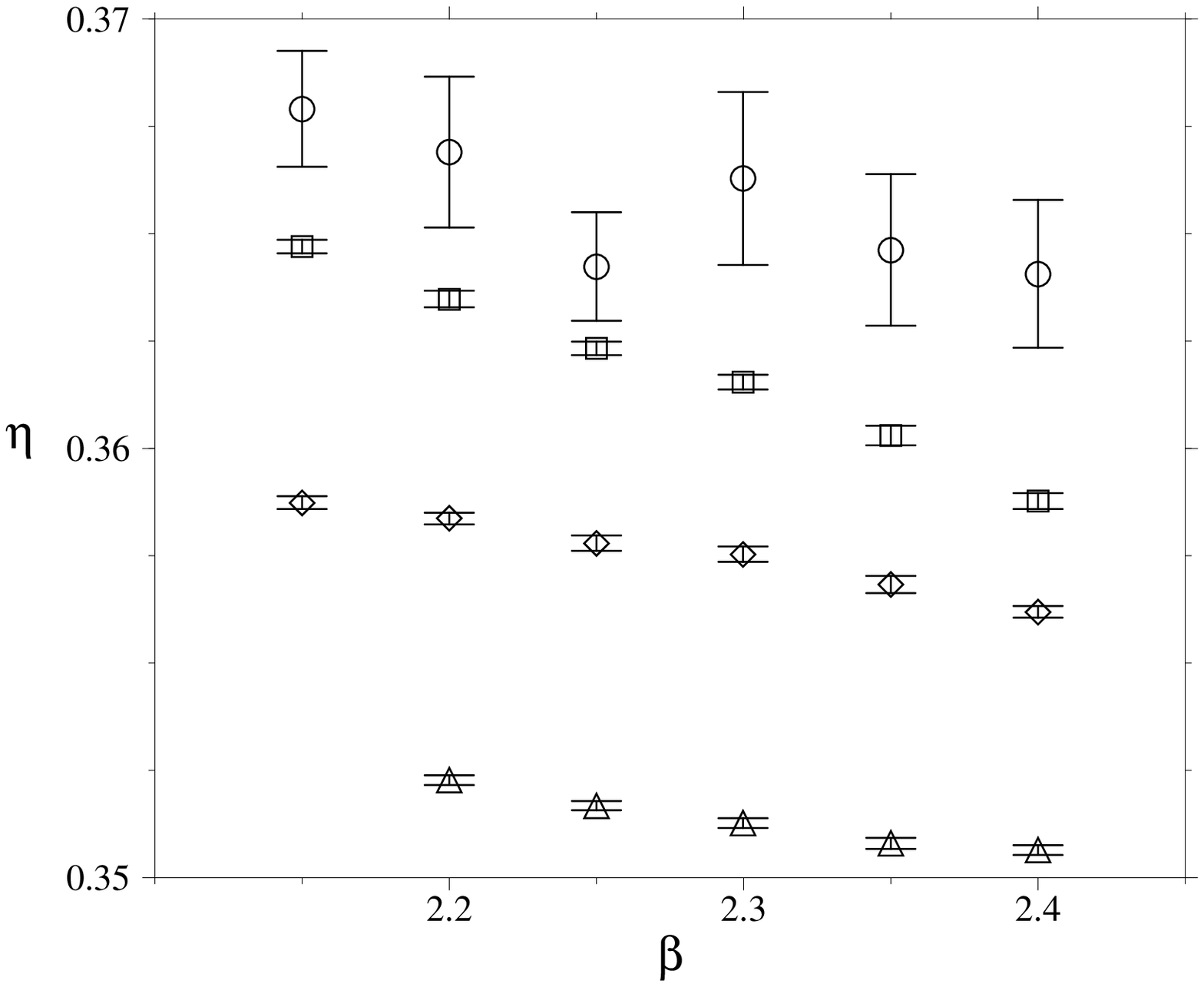,width=0.5\textwidth,silent=}}

\centerline{Fig.~8}
\end{figure}

To monitor the quality of each algorithm we consider 
\beq
\eta ~=~ \frac{1}{2}\,(1-\,\langle\vec{n}_{i,\alpha}\cdot \vec{n}_{j,\beta}\rangle\,)\,,
\eeq
where $\vec{n}_{i,\alpha}\cdot\vec{n}_{j,\beta}$ is the gauge invariant
scalar product of two neighboring vectors. 
The dependence $\eta(\beta)$ on $12^4$ lattice is presented on Fig.~8,
where circles denote the local minimization without any preconditioning,
while the local minimization with minimal Landau and Maximal Abelian gauge preconditioning
are the squares and diamonds, respectively. Triangles are the results for the
simulated annealing algorithm.

It should be noted that although the absolute values of $\eta$ varies slightly for
various algorithms, the corresponding monopole densities differ by an order of magnitude.
Note also that in the case of Maximal Abelian gauge preconditioning there was in fact no
local minimization at all (see section~4). Nevertheless the $\eta$ values
and monopole densities with MAA gauge are at least comparable with ones obtained
via the simulated annealing.

Let us now turn to the physically interesting question about the monopole density.
Fig.~9 represents the densities of artifacts (circles) and monopoles (squares)
versus bare coupling constant $\beta$ on a $14^4$ lattice. In view of the above
discussion of the monopole-artifact mixing it should not be a surprise that none
of the graphs on Fig.~9 follows the renormalization group prediction
\beq
\label{renorm}
\rho_{lat} ~=~ \frac{\rho_{phys}}{4\Lambda^3}\,
\left[\frac{6\pi^2}{11}\beta\right]^{153/121}
\,\exp\{ - \frac{9\pi^2}{11}\beta\}\,,
\eeq
which is to be valid for physical monopoles. Moreover, Fig.~9 clearly confirms
the actionless nature of artifacts since their density is almost constant in the
whole range of $\beta$ considered.

\begin{figure}[ht]
\centerline{
\psfig{file=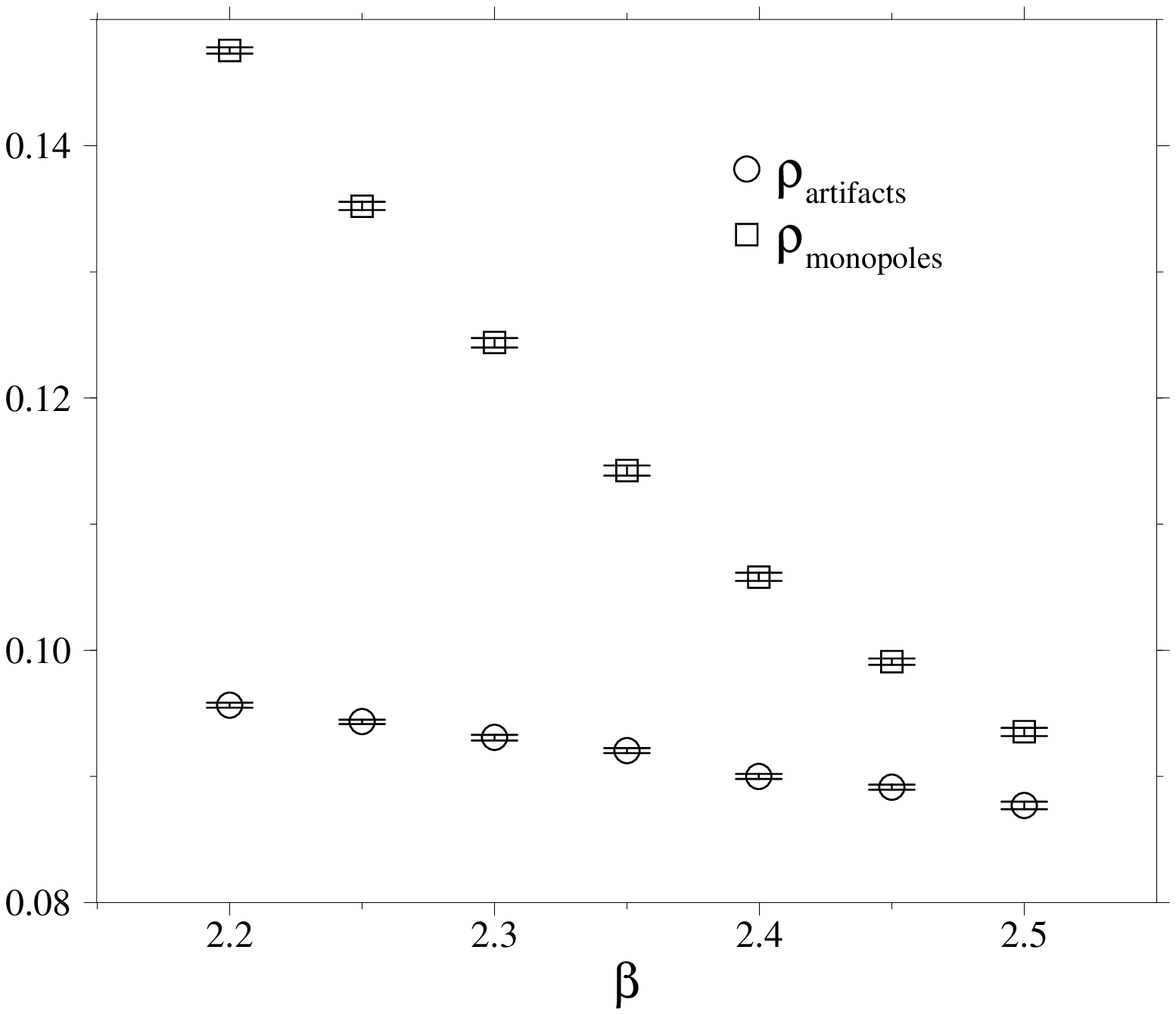,width=0.49\textwidth,silent=}
\psfig{file=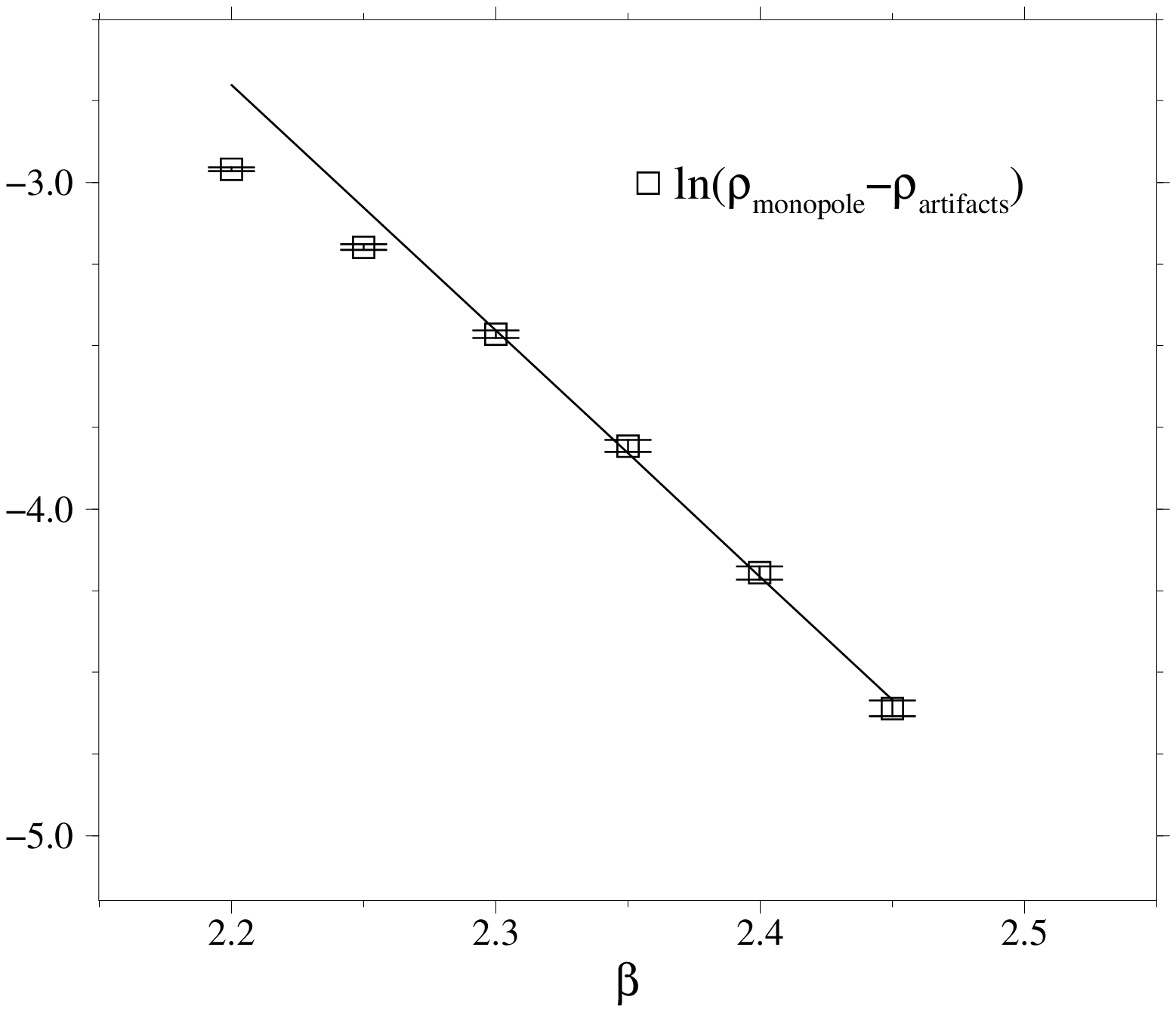,width=0.5\textwidth,silent=}
}

\centerline{Fig.~9~~~~~~~~~~~~~~~~~~~~~~~~~~~~~~~~~~~~~~~~~Fig.~10}
\end{figure}

As we have already discussed, the monopole-artifact separation is {\it a priori} not
a well defined procedure at a finite lattice spacing. Nevertheless, it is amusing to
note that in case of $D=4$ $SU(2)$ lattice gluodynamics the difference between the 
monopole
and artifact densities follows Eq.~(\ref{renorm}) sharply, see Fig.~10.
It seems so that in four dimensions and in the weak coupling regime artifacts do indeed
form small dipoles with the monopoles living on the dual lattice.
In this way one can explain the observed scaling behavior of
$\rho_{phys} = \rho_{monopoles} - \rho_{artifacts}$
although there are strictly speaking no rigorous arguments why $\rho_{phys}$ scales.
Indeed, the monopole-artifact dipole picture is just an unproved assumption
which, however, works unexpectedly well in $D=4$ $SU(2)$ LGT. The rigorous
artifact separation is still an unresolved problem which is beyond the scope
of the present publication.

Assuming that the above dipole picture is valid one can estimate the density
of the physical monopoles in the continuum limit. As follows from Fig.~10,
$\ln\rho_{phys}/4\Lambda^3 \approx 11.9$ and using the numerical value of the string
tension $a\sqrt{\sigma} = 0.1326$ at $\beta = 2.6$ (see, e.g. Ref~\cite{Teper}) one gets:
\beq
\rho_{phys}~=~ (1.72 \, \sqrt{\sigma})^3 ~\approx~ (760 \mbox{ MeV })^3\,,
\eeq
where the conventional value $\sqrt{\sigma}=440\mbox{ MeV }$ has been used.

%====================================================================
\subsection*{Conclusions}
\noindent
We have implemented the gauge invariant monopole charge definition in the $SU(2)$
lattice gluodynamics. The lattice implementation is of particular importance
since there are no analytical approaches to the gauge invariant monopoles
in the continuum limit. Indeed, the corresponding classical Wu-Yang solution
is unstable.
Our lattice construction is based on the  continuum definition
of the monopole charge proposed recently and in fact follows it as close as possible.
In this way we were able to obtain integer valued conserved monopole
currents. 

As might be expected, the lattice formulation brings in
 a set of specific problems. One of them is the appearance
of the effective Ising model. In fact,  the gauge invariance of monopoles on the
lattice crucially depends on our ability to find its ground state,
which is a non-trivial task because the model is frustrated. But
at least in principle this problem might be solved. Moreover, here we found
a hint on the distinguished role of the Maximal Abelian gauge, which proved to be
successful in the context of Abelian monopole confinement mechanism.

Another problem specific for the lattice is the emergence of the lattice artifacts.
In more detail:

{\it i)} Artifacts are mixed up with physical monopoles. Thus it is necessary
to subtract their contribution in order to get physical results.
For simple quantities, like monopole density 
the artifact separation is indeed possible. Then the density of
physical monopoles seems to scale correctly towards the continuum limit.
However, it is not yet completely clear how to get rid of the
artifacts in general case.

{\it ii)} Due to artifacts mixing the hypercubical geometry of the original lattice
has to be changed. The monopole currents are still conserved, but the conservation
takes place on a complicated lattice, the geometry of which is still not well understood.
Probably the understanding of the lattice geometry
will help to treat artifacts in a more rigorous way.

%====================================================================
\subsection*{Acknowledgments.}
\noindent
We acknowledge thankfully the fruitful discussions with V.~Belavin, M.N.~Chernodub,
 R.~Hofmann, M.I.~Polikarpov,
L.~Stodolsky and V.I.~Zakharov. The work was partially supported by grants
RFFI-02-02-17308, RFFI-0015-96786, INTAS-00-0011 and CRDF award RP1-2364-MO-02.

%==========================================================================================

\end{document}